\documentclass[12pt]{article}
\usepackage{epsfig}
\usepackage{latexsym}
\newcommand{\be}{\begin{equation}}
\newcommand{\ee}{\end{equation}}
\newcommand{\bea}{\begin{eqnarray}}
\newcommand{\eea}{\end{eqnarray}}

\newcommand{\mpl}{M_{\rm pl}}
\newcommand{\lpl}{L_{\rm pl}}

\newcommand{\vk}{{\bf k}}
\newcommand{\vx}{{\bf x}}
\newcommand{\vy}{{\bf y}}

\newcommand{\ren}{{\rm ren}}
\newcommand{\erf}{{\mathrm{erf}}}
\newcommand{\et}{{\rm e}}

\begin{document}
\baselineskip=20pt

\begin{flushright}
hep-th/0410270 \\ EFI-2004-33  \end{flushright}

\vspace*{2cm}

\begin{center}
{\Large{\bf Spontaneous Inflation and the Origin\\
of the Arrow of Time}}

\vspace*{0.3in}
Sean M.\ Carroll and Jennifer Chen
\vspace*{0.3in}

\it Enrico Fermi Institute, Department of Physics, and\\
Kavli Institute for Cosmological Physics,
University of Chicago\\
5640 S.~Ellis Avenue, Chicago, IL~60637, USA\\
{\tt carroll@theory.uchicago.edu, jennie@theory.uchicago.edu} \\
\vspace*{0.2in}
\end{center}

\begin{abstract}
We suggest that spontaneous eternal inflation can provide a natural
explanation for the thermodynamic arrow of time, and discuss the
underlying assumptions and consequences of this view.  In the absence
of inflation, we argue that systems coupled to gravity usually evolve
asymptotically to the vacuum, which is the only natural state in a
thermodynamic sense.  In the presence of a small positive vacuum
energy and an appropriate inflaton field, the de~Sitter vacuum is
unstable to the spontaneous onset of inflation at a higher energy
scale.  Starting from de~Sitter, inflation can increase the total
entropy of the universe without bound, creating universes similar to
ours in the process.  An important consequence of this picture is that
inflation occurs asymptotically both forwards and backwards in time,
implying a universe that is (statistically) time-symmetric on
ultra-large scales.
\end{abstract}

\vfill
\newpage
\tableofcontents

\newpage
\baselineskip=14pt

\section{Introduction}

The role of initial conditions in cosmology is unique within
the physical sciences.  We only have a single observable universe,
rather than the ability to change boundary conditions and run
experiments multiple times.  A complete theory of cosmology therefore
involves not only a set of dynamical laws, but a specification of the
particular initial conditions giving rise to the universe we see.

One could certainly argue that the origin of our initial conditions is
not an answerable scientific question.  Given the state of our universe
at the present time, and a complete set of dynamical laws describing
its evolution, we can in principle solve for the entire
history, including whatever the initial state was.  Ultimately, we are
stuck with the boundary conditions we have.  Similarly, however, we are
stuck with the laws of physics that we have, but this constraint doesn't
stop us from searching for deep principles underlying their nature.
It therefore seems sensible to treat our initial conditions in the
same way, and try to understand why we have these conditions rather than
some others.

In particular, we would like to know whether the initial conditions for our
universe are in some sense ``natural.''  The notion of naturalness in
this context is not precisely defined, but in simple cases we know it
when we see it.  Our current universe exhibits a pronounced
matter/antimatter asymmetry, which seems unnatural, so we search for
mechanisms of baryogenesis to create this asymmetry dynamically.  The
large-scale homogeneity and isotropy of our universe is unstable, so
we search for mechanisms in the early universe to generate this smoothness
dynamically.

One aspect of the boundary conditions for our universe that remains
puzzling, or at least controversial, is the arrow of time (for discussions
see \cite{price,albrecht03,Coule:2002zb}).  The
Second Law of Thermodynamics tells us that the entropy of a closed
system is non-decreasing as a function of time.  A comoving patch
corresponding to our observable universe is not precisely a closed
system, but is pretty close, since homogeneity of the universe near
the boundary implies there should be no net entropy flux into or out
of our patch.  And indeed, we observe increasing entropy in our
universe; the thermodynamic arrow of time is the direction picked out
by this increase of entropy.

Long ago Boltzmann understood the increase of entropy
as a statistical phenomenon, as evolution from a very small region of
phase space to a larger region.  The origin of the Second Law is
traced back to initial conditions:  the early universe had an extremely
low entropy, allowing it to continue to increase thereafter.  Even
after 13.7 billion years of entropy growth, it remains low in our
current universe.  	Within our observable patch, the entropy in ordinary
matter is of order
\be
  S_M(U) \sim 10^{88}\ .
  \label{meu}
\ee
This figure is dominated by photons and neutrinos; dark matter may also
contribute, but wouldn't significantly change the final answer.  However,
Penrose has emphasized \cite{penrose} that this entropy could be much
larger.
Our universe remains quite smooth on large scales, which is a finely-tuned
condition to be in.  To see this, we can simply compare the current
entropy to what it could have been, for example if more energy were
stored in black holes.
The Bekenstein-Hawking entropy of a black hole \cite{Bekenstein:1975tw,
Hawking:1974sw} is proportional to its
horizon area,
\be
  S_{BH} = {A\over 4G}\ .
\ee
We believe that galaxies similar to ours contain supermassive black
holes at their centers, of order a million solar masses; the
entropy of a single such black hole is
\be
  S_{BH} \sim 10^{89}\left({M_{BH}\over 10^6 M_\odot}\right)^2\ .
\ee
Any one such black hole therefore contains more entropy than all of the
ordinary matter in the visible universe.  But there are
perhaps $10^{10}$ such black holes in the universe; their total entropy is
thus
\be
  S_{BH}(U) \sim 10^{99}\ ,
  \label{bheu}
\ee
which represents most of the entropy within our observable universe.
This seems like a large number, but should be compared to how large it
could be.  The total amount of mass in the observable universe is
\be
  M_U \sim \rho H_0^{-3}\sim G^{-1}H_0^{-1} \sim 10^{22} M_\odot\ .
\ee
If all of this mass were collected into a single black hole, the entropy
would be
\be
  S_{\rm max}(U) \sim 10^{121}\ ,
  \label{eumax}
\ee
larger than the actual entropy by a factor of $10^{22}$.  Moreover,
in the early universe there weren't any black holes, and the entropy
was dominated by the matter entropy (\ref{meu}), smaller than the
equivalent black-hole entropy by a factor of $10^{33}$.  So the entropy
of our universe seems to be very small, and seems to have evolved from
an era where it was significantly smaller.

The initial conditions for our observable universe thus have extremely
low entropy.  We would like to understand these conditions as arising
through dynamical evolution from a natural starting point; it is
not immediately obvious how this could be achieved, however, since
dynamical evolution tends to increase the entropy.

The low entropy of our current universe is clearly related to its
homogeneity and isotropy, features which are supposed to be explained
by the inflationary universe scenario \cite{Guth:1980zm,Linde:1981mu,
Albrecht:1982wi,Starobinsky:1979ty,Starobinsky:1980te}.
It is therefore tempting to invoke inflation as the origin of the
thermodynamic arrow of time, and indeed this move has been made
\cite{Davies:1983nf}. The converse argument, however, has also been advanced:
that inflation never works to decrease the entropy, and posits that
our observable universe originates in a small patch whose entropy is
fantastically less than it could have been, so that in fact there is
a hidden fine-tuning of initial conditions implicit in the inflationary
scenario, which consequently (in the extreme version of this reasoning)
doesn't explain anything at all \cite{penrose,Page:1983uh,unruh,
Hollands:2002yb}.

In this paper we suggest a resolution of this apparent
tension between inflation and the Second Law.  We first discuss in
general terms what sort of evolution for the universe would qualify
as naturally giving rise to an arrow of time through dynamical
processes.  We then review the arguments concerning the role
of inflation in setting up the early universe in a low-entropy state,
concluding that ordinary inflation does not directly explain the origin
of the arrow of time if one believes in unitary evolution.
We then argue that, in the absence of a vacuum
energy or inflation of any sort, systems coupled to gravity generically
evolve to flat, empty space.  In the presence of a small positive
vacuum energy (such as seems to exist in our universe
\cite{Riess:1998cb,Perlmutter:1998np,Carroll:2000fy}), systems
will tend to initially empty out to an approximate de~Sitter vacuum.
If there is also at least one scalar inflaton field with an appropriate
potential, however, there is an instability:
quantum fluctuations in this field will eventually
create an inflating patch, and according to the traditional
arguments for eternal inflation \cite{Vilenkin:1983xq,Linde:1986fc,
Linde:1986fd,Goncharov:1987ir,Guth:2000ka} the total physical volume of
inflating space will continue to increase thereafter, while
``pocket universes'' drop out of the inflating background and reheat.
The idea of inflation beginning spontaneously 
has been suggested previously 
\cite{Vilenkin:1982de,Vilenkin:1983xq,Linde:1983mx,Vilenkin:1984wp,
Starobinsky:1986fx,Goncharov:1986ua,
Farhi:1986ty,Vilenkin:1987kf,Farhi:1989yr,Fischler:1989se,Fischler:1990pk,
Linde:1991sk}.  Recently, Garriga and Vilenkin \cite{Garriga:1997ef}
have suggested that spontaneous inflation in a low-energy de~Sitter
background could lead to a recycling universe, while Dyson, Kleban
and Susskind and Albrecht and Sorbo have examined spontaneous inflation
in the context of a causal-patch view inspired by
the holographic principle \cite{Dyson:2002pf,
Albrecht:2004ke}; we discuss the relationship between our
picture and these ideas in greater detail below.

The crucial features of this scenario are thus the following:
\begin{enumerate}
\item Generic initial states empty out and approach de~Sitter space.
\item In the presence of an inflaton field, the de~Sitter vacuum will be
unstable to the onset of eternal inflation.
Once eternal inflation starts, the entropy of the
universe can grow without bound, never reaching thermal equilibrium.
\item In the process of increasing the
entropy, eternal inflation creates large regions of space similar to
our observable universe.
\end{enumerate}
As an important consequence of our reasoning,
we obtain a picture of the ultra-large-scale structure of the universe.
As in any model of eternal inflation, the ultimate state of the universe is a
fractal distribution of inflating and post-inflating regions
\cite{Linde:1986fd,Goncharov:1987ir,Aryal:1987vn,Winitzki:2001np},
in which our local Big Bang is not a particularly unique event.
We also predict that this structure should be recovered infinitely
far into the past, with a {\it reversed} thermodynamic arrow of time.  Our
overall universe is therefore statistically time-symmetric about some
Cauchy surface of minimum entropy.  The details of the state on this
surface are not important; starting from a generic state, inflation
eventually occurs towards both the past and future, erasing any
information about the ``initial'' conditions.

\section{Explaining the Arrow of Time Dynamically}
\label{arrow}

Our goal is to understand how the arrow of time might arise from
natural initial conditions for the universe.  In this section we
discuss some necessary features of any purported resolution of this
issue.

To make any progress we need to have some understanding of what
it means for an initial condition to be ``natural.''  If an unknown
principle of physics demands specific initial conditions (but not
final ones), there is no problem to be solved; it may simply be that
the universe began in a low-entropy initial state and has
been evolving normally ever since.  This might be the case, for example,
if the universe were created ``from nothing,'' such that the initial
state was {\it a priori} different from the final state 
\cite{Vilenkin:1982de,Linde:1983mx,Hartle:1983ai,Vilenkin:1984wp}, 
or if Penrose's explicitly time-asymmetric Weyl Curvature
Hypothesis were true \cite{penrose}.  There is, of course, no
way to rule out this possibility.

Nevertheless, it would be more satisfying if we
could somehow understand our apparently low-entropy initial condition
as an outcome of dynamical evolution from a generic state.
In particular, we
wish to avoid any explicit violation of time-reversal symmetry in
the specification of the initial condition.  The very word ``initial,''
of course, entails a violation of this symmetry; we should therefore
apply analogous standards to the final conditions of the universe.
Price \cite{price} has proposed a ``double-standard principle'' -- anything
that is purportedly natural about an initial condition for the universe
is equally natural when applied to a final condition.  The conventional
big-bang model clearly invokes unnatural initial conditions, as we
would not expect the end state of a recollapsing universe to arrange
itself into an extremely homogeneous configuration (as we discuss in
more detail below).

Attempts to comply with the double-standard principle (or its
underlying philosophy) have occasionally led to proposals for
explicitly time-symmetric cosmologies in which entropy is low both
at the beginning and the end of time.  In 1962, Gold proposed a
recollapsing universe with similar conditions imposed at the Bang
and Crunch \cite{gold,price}.
In the Gold universe, entropy increases as the
universe expands, reaches a maximum, and then begins to decrease
again.  Hawking at one time suggested that such behavior could follow
naturally from quantum cosmology, but Page argued persuasively
that it was not necessary \cite{Hawking:1985af,Page:1985ei}.
More recently, Gell-Mann and Hartle have investigated the
possibility of imposing quantum-mechanical boundary conditions in both
the past and future \cite{Gell-Mann:1991ck}.  It might be difficult to find a
consistent implementation of the Gold universe, and Hartle and
Gell-Mann even suggest that the possibility of a time-symmetric
Bang and Crunch may be ruled out by observation (from the absence of
observed backward-moving radiation emitted by stars in the future).
For most people, the important objection to these models is
simply that they seem hopelessly {\it ad hoc}; imposing boundary
conditions both in the past and the future would appear to be
the opposite of ``natural.''  Price \cite{price} argues that imposing
low-entropy conditions in the past and the future is the only way
to avoid conflict with the double-standard principle.  We will
suggest a loophole in this argument, by suggesting that {\it high}-entropy
conditions obtain in the far past and far future, and reconciling
this unusual claim about the past by arguing that our
observed early universe is merely a small part of a larger state.

One way to understand the concept of naturalness would be to imagine that
we had a measure on the space of microstates describing the universe,
and an appropriate way of coarse-graining into macroscopically
indistinguishable states.  The entropy is then simply the logarithm
of the measure on any collection of macroscopically indistinguishable
microstates, and natural initial conditions are those with large entropy.
Unfortunately, we do not have
a reliable understanding of entropy in the context of quantum
gravity; the fundamental degrees of freedom themselves remain unclear.
Even without a quantitative understanding at the level of statistical
mechanics, however, there are certain essential features that entropy
must have, which will be sufficient to highlight problems with the
conventional picture.

In ordinary thermodynamics, the fundamental property of the entropy is
that it tends to increase (or at least not decrease) with time.
Of course, this property is understood as a consequence of low-entropy
initial conditions.  If we imagine choosing initial conditions
randomly, with uniform probability for choosing any microstate,
a system with a finite number of degrees of freedom will
typically be found in a state of maximum entropy.  Maximum-entropy
states represent thermal equilibrium, and are generally static
(at least macroscopically).  Statistical fluctuations
will allow the entropy to occasionally, although rarely, decrease from
that point, and then return to equilibrium.
In thinking about the
early universe, it is worth keeping in mind that a state that rapidly
evolves into something else cannot possibly be a ``natural'' state in
the sense of having maximal entropy, since natural states should be
static (at least statistically).

Given an infinite
amount of time, a finite system will approach arbitrarily close to
all possible configurations consistent with the existence of certain
conserved quantities (Poincar\'e recurrence).  It is therefore irresistible
to ask whether our observed arrow of time could arise as the result of
such a fluctuation.  The answer seems to be ``no,'' for the basic reason that
our observed universe has a much lower entropy than can be explained in
this way.  For example, if we appeal to the anthropic principle to
understand our current state, we would only require a fluctuation in
entropy large enough to allow for a single conscious observer, not an
entire universe (``Boltzmann's Brain'' \cite{bt,albrecht03,Albrecht:2004ke}).  
Further, even if we
consider the set of universes which are similar to ours, most of them
are not thermodynamically sensible, in the sense of arising from
lower-entropy initial conditions \cite{Dyson:2002pf}.
It therefore requires more than
a simple fluctuation out of thermal equilibrium to explain our current
state.  (In the context of causal-patch physics this problem becomes
particularly acute, as there really are only a finite number of 
degrees of freedom to be considered.  Albrecht and Sorbo have 
argued that inflation can overcome the Boltzmann's brain paradox
even in this context \cite{Albrecht:2004ke};
as we discuss in Section~\ref{causalpatch}, we are skeptical that
any theory of fluctuations around equilibrium would be able to
resolve this puzzle.)

We are left with the following conundrum:  we would like to explain
our currently observed universe as arising from natural initial
conditions, but natural means high-entropy, and high-entropy implies
equilibrium configurations with occasional fluctuations, but not ones
sufficient to explain our observed universe.  However, there is
one loophole in this reasoning, namely the assumption that there is
such a thing as a state of maximal entropy.  If the universe truly
has an infinite number of degrees of freedom, and can evolve in a
direction of increasing entropy from any specified state, then an
explanation for the observed arrow of time arises more naturally.

\begin{figure}[t]
\centerline{
\psfig{figure=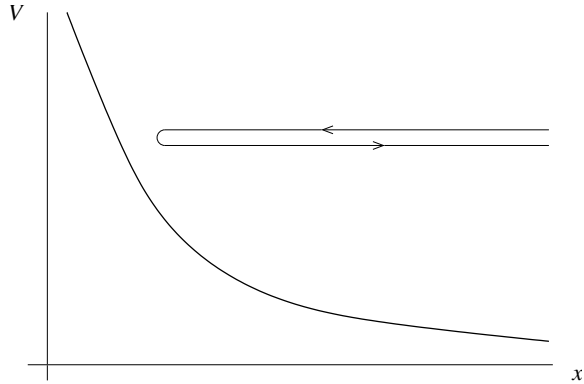,height=5cm}}
\caption{Motion of a particle in a potential that rolls to infinity
without ever reaching a minimum.  All trajectories in this potential
come in from infinity, reach a turning point, and return to infinity.}
\label{inandout}
\end{figure}

The situation is analogous (although not equivalent) to the behavior
of a particle moving in a potential that rolls off to infinity without
having any minima, such as $V(x)=1/x$, as portrayed in
Figure~\ref{inandout}.  A free particle in such a potential comes in
from infinity, reaches a turning point, and returns to infinity.
On any such trajectory, every point except the turning point is either
moving towards, or coming from, larger values of $x$.  Now imagine
that entropy in the universe behaves the same way.  It would not
be surprising to find ourselves in a situation where the entropy
were evolving; it would be almost inevitable, and certainly perfectly
natural.  Indeed, we would simply {\it define} the ``past'' in such
a universe to be the direction of time in which the entropy was
{\it locally} decreasing.  All that is needed to have an arrow of
time arise dynamically is for the entropy to be unbounded above, so
that it can always increase from any given starting point.

\begin{figure}[t]
\vskip-0cm
\centerline{
\psfig{figure=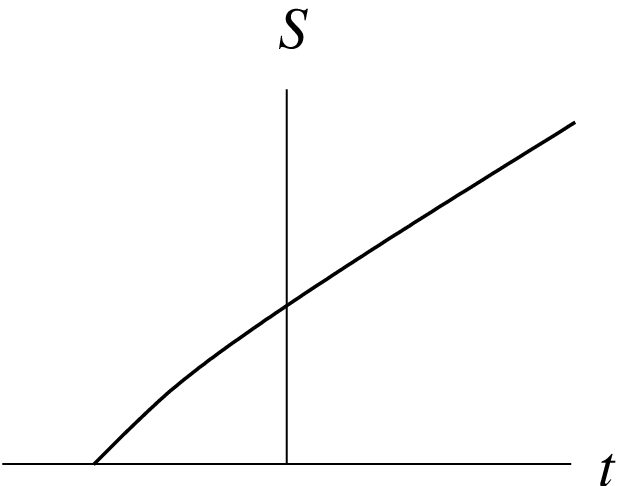,height=5cm}$\qquad$ \psfig{figure=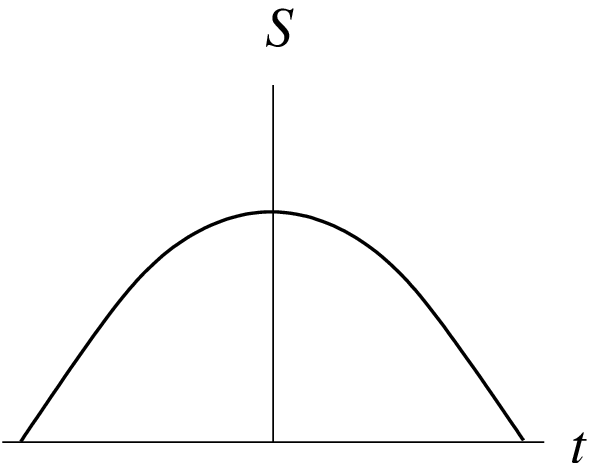,height=5cm}}\centerline{\psfig{figure=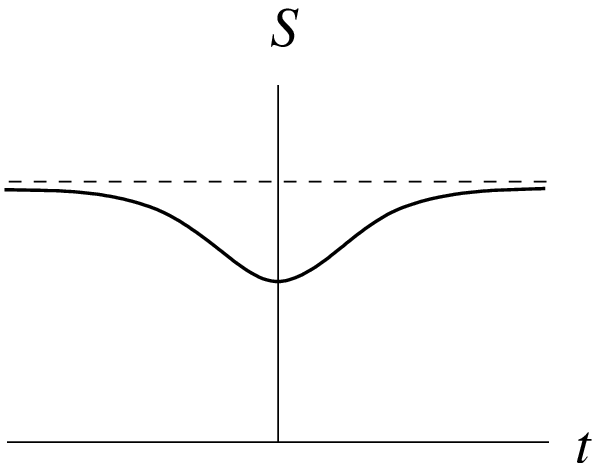,height=5cm}$\qquad$ \psfig{figure=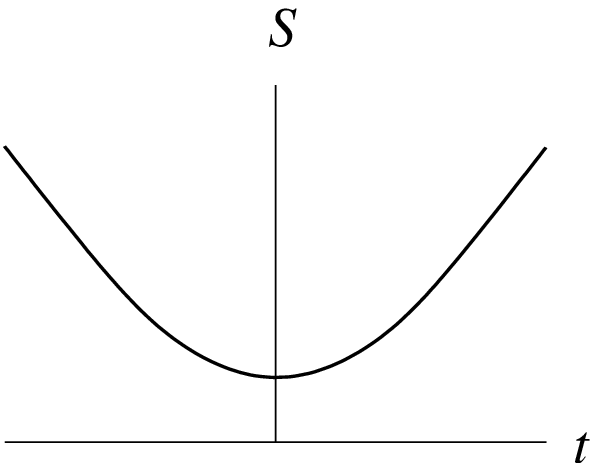,height=5cm}}
\caption{Possible evolutions of the entropy of the universe.  At
top left, a universe that comes into existence
with prescribed low-entropy conditions in the
past; at top right, a universe with low-entropy conditions in both
the past and future.  At bottom left, a fluctuation from an equilibrium
value of maximum entropy, portrayed by the dashed line.  At bottom
right, a universe with unbounded entropy to the past and future.}
\label{entropychoices}
\end{figure}

In Figure~\ref{entropychoices} we portray the various possible ways
to address the arrow-of-time problem, in terms of the evolution of
entropy as a function of time.  The first choice is to impose
time asymmetry by hand, simply insisting on low-entropy initial
conditions; this scenario is plausible, but we would prefer not to rely
on an {\it ad hoc} initial condition.  The second is to impose
low-entropy conditions at both initial and final times, as in the Gold
universe.  Such a picture is also possible, but equally {\it ad hoc},
as it implies that the universe is actually finely-tuned at every
possible moment (to be consistent with the future boundary condition).
The third possibility is to imagine that the early universe is a
random low-entropy fluctuation from an equilibrium state; this seems
implausible on anthropic grounds.  The final possibility, advocated
here, is that there is no equilibrium configuration for the universe;
instead, the entropy can increase without bound, and will generally
do so from any given boundary condition.  (Such a boundary condition
is not actually imposed at a boundary, but at some arbitrary moment.)
We should stress that there might also be, and perhaps should be,
some procedure for actually specifying this boundary condition; but
according to our picture this specification will leave no observable
imprint on our universe, and in that sense is truly irrelevant.

Of course, it is certainly not {\it sufficient} to imagine that the
entropy of the universe is unbounded, although it seems to be
necessary.  We also need to understand why the process of entropy
creation would create regions of spacetime resembling our observable
universe.  This is where we can sensibly appeal to inflation, as
discussed in the next section.

\section{Inflation and its Discontents}
\label{critique}

Inflation is an extremely powerful and robust idea.  It purports
to solve several severe fine-tuning problems of conventional
cosmology, including the horizon, flatness, and monopole problems.
As a bonus, it provides a natural mechanism for the origin of the
approximately scale-free density perturbations that serve as the
origin of structure in our universe.  Perhaps equally importantly,
it provides a natural explanation for why the observable universe should be
large and expanding at all, as inflation can create a fantastic number
of particles from a microscopically small initial patch of spacetime.

However, there are some lurking problems in our understanding of the
onset of inflation and the role of initial conditions.  In this
section we review some of the existing controversies over
initial conditions for inflation and the role of entropy, arguing
as effectively as we can that there really is a serious issue that has yet
to be addressed concerning the onset of inflation.  This discussion will
set the stage for our suggested resolution of this problem in later
sections.

\subsection{Starting Inflation}

Let's review the standard picture of inflation, choosing for definiteness
a simple quadratic potential
\be
  V(\phi) = {1\over 2}m^2 \phi^2\ ,
  \label{potential}
\ee
where the scalar field $\phi$ is the inflaton, and, in order to get the
correct amplitude of density perturbations, the mass parameter
is chosen to be
\be
  m \sim 10^{13}~{\rm GeV}\ .
  \label{mass}
\ee
The Friedmann equation can be written
\be
  H^2 = {8\pi\over 3\mpl^2}\rho\ ,
  \label{feq}
\ee
where the ordinary Planck mass is $\mpl = 1/\sqrt{G} \sim 10^{19}$~GeV.
In this model, inflation can occur when the field value is above
a certain critical value
\be
  \phi_I \sim \mpl\ ,
  \label{phii}
\ee
so that the energy density $\rho_I$ is near the scale of grand unification
$\rho_I \sim (10^{16}~{\rm GeV})^4$, and the Hubble radius about
$10^6$ times the Planck length,
\be
  H_I^{-1} \sim (10^{13}~{\rm GeV})^{-1} \sim 10^6 \lpl\ .
  \label{hi}
\ee
(This is the requirement for inflation with an appropriate amplitude
of density fluctuations; eternal inflation only occurs at a higher
scale, as we discuss later.)
In order for inflation to begin, there must be a region that is smooth,
dominated by potential energy, and of radius $L_I$ larger than $H_I^{-1}$
\cite{Linde:1985ub,Albrecht:1986pi,Goldwirth:1991rj,Vachaspati:1998dy}.  If
the initial value of the scalar field is sufficiently large, inflation
will occur for more than the sixty $e$-folds of accelerated
expansion necessary to bring our entire currently observable universe
into causal contact.  At the end of the inflating period, we imagine that
the energy of the inflaton is converted by reheating into excitations
of ordinary matter and radiation.

During the period of accelerated expansion, the spatial curvature of
the initial patch is driven to zero, and any small inhomogeneities
diminish as the universe rapidly stretches.  In this way, a very small
proto-inflationary region can naturally evolve into our observable
patch today, explaining its high degree of homogeneity and isotropy.
Thus, although smoothness and flatness are unstable features in
conventional
cosmology, inflation drives the universe towards such a state, thereby
(so the story goes) explaining these otherwise puzzling features.

Inflation thus claims to provide a mechanism by which a universe like
ours can arise robustly from random initial conditions, not requiring
fine-tuning to preserve homogeneity and isotropy.  As mentioned in the
introduction, a simple alternative would simply be that the universe did
not start with random initial conditions, but rather with the appropriate
ones to evolve straightforwardly into our present state.  In the absence
of some specific theory of initial conditions that would actually
predict such an initial state, this latter point of view doesn't actually
explain anything; inflation, in contrast, purports to explain why our
universe should look the way it does almost independently of the actual
initial conditions.  However, the strength of this explanation relies
heavily on a certain intuitive notion of what constitutes a ``random''
or ``generic'' state.  In particular, there is an assumption
that it is more likely to find an appropriate proto-inflationary patch
than it is to find a patch resembling an early stage of the conventional
hot Big Bang model.

Some specific arguments have been advanced that it cannot be {\em too}
unlikely to find a patch of
space with appropriate proto-inflationary conditions lurking within
a randomly-chosen initial state.  Consider, for example, the
allowed modes of a field confined to a small patch of linear size
$L_I$.  Since this patch is only slightly larger than the Planck length,
there are only a certain number of modes with sub-Planckian energies
[perhaps $(L_I/\lpl)^3 \sim 10^{18}$]; we need only to have most of
these modes be in their ground states in order for inflation to begin.
Another version of the argument comes from
Linde's chaotic inflationary scenario \cite{Linde:1983gd}, in which
randomly fluctuating conditions in the early universe will occasionally
give rise to an appropriate patch that begins to inflate; once inflation
does start, the fantastic increase in the volume of space guarantees
that most of the universe will eventually be found in the inflationary
or post-inflationary regions.
Indeed, inflation could plausibly begin in a region
of a single Planck volume with an energy density at the Planck scale;
all we need is for a fluctuation to give rise to one such region
\cite{Kofman:2002cj}.

\subsection{The Unitarity Critique}

A critique of the conventional inflationary scenario
has been advanced, in slightly different
versions, by Penrose \cite{penrose}, Unruh \cite{unruh}, and
Hollands and Wald \cite{Hollands:2002yb}, and in the context of causal-patch
physics by Dyson, Kleban and Susskind \cite{Dyson:2002pf}.  The claim of these
counterarguments is essentially that it is extremely unlikely to find a
patch of space in the appropriate proto-inflationary conditions --
less likely, even, than to find the universe in the initial conditions
for the conventional Big Bang model.  We will first present this argument in
two similar forms:  one using entropy as a
measure on the space of initial conditions, and one appealing to our
notions of what would be likely for a collapsing universe.  We will
then discuss the closely related issues raised by the causal-patch
picture suggested by holography, which implies that one should consider
only finite number of degrees of freedom contained in each de~Sitter
volume.

\subsubsection{Entropy version}

The most straightforward and dramatic version of this argument
involves comparing the entropy of the early universe to that of the
universe today \cite{penrose}.  The entropy is the logarithm of the
volume corresponding
to macroscopically indistinguishable states, and therefore measures
the likelihood of ``randomly'' choosing a certain condition, if
we assign uniform measure to each microscopic configuration.  (This
is, of course, a nontrivial assumption. In particular, a universe
that fluctuates into existence out of nothing may do so into a very
specific state, not a randomly-chosen one.)  The
crucial point here is that our currently observable universe and the
small patch of early universe that grows into it are not two different
physical systems; they are two different {\em configurations} of the
{\em same system}.

This fact seems at once completely obvious and deeply counterintuitive.
Of course the small patch in the early universe and the currently
observable patch into which it grows are two configurations of the same
system; after all, one evolves into the other.  On the other hand, they
appear very different -- different volumes, different total energies,
different numbers of particles.  It is therefore tempting to judge them
according to different sets of standards, while in fact they simply
represent different subsets of the state space of (the comoving
volume corresponding to) our universe.
This is a consequence of the fundamental weirdness of statistical
mechanics in the presence of gravity, about which we will have more to
say in the next section.

Let us then consider the entropy of the proto-inflationary
patch.  We don't have a reliable way to calculate this entropy, but
there are reasonable guesses, and the precise answer is not important.
Since the patch is taken to be in a quasi-de~Sitter initial state, we
could, for example, consider the corresponding de~Sitter entropy for
a patch of that size,
\be
  S_I \sim \left({L_I \over \lpl}\right)^2 \sim 10^{12}\ .
\ee
Alternatively, we could simply count the number of Planck volumes in
the proto-inflationary region, $(L_I/\lpl)^3 \sim 10^{18}$, imagining
that there is of order one degree of freedom per Planck volume.  Of course,
the proper measure of the entropy might be infinite, if there are
truly an infinite number of degrees of freedom; but in that
case the entropy of the late universe is also infinite, and the numbers
given here are an appropriate standard for comparison.

The point of the entropy argument is then very simple:  the entropy of the
patch that begins to inflate and expands into our observable universe
is far less than our current entropy (\ref{bheu}), or even than
the entropy at earlier stages in our comoving volume, given approximately
by the matter entropy (\ref{meu}).  {}From the point of view of the Second
Law, this makes sense; the entropy has been increasing since the onset
of inflation.  But from the point of view of a theory of initial
conditions, it strongly undermines the idea that inflation can naturally
arise from a random fluctuation.  In conventional thermodynamics,
random fluctuations can certainly occur, but they occur with exponentially
smaller likelihood into configurations with lower entropy.
Therefore, if we are imagining that the conditions in our universe
are randomly chosen, {\em it is much easier to choose
our current universe than to choose a small proto-inflationary patch.}
The low entropy of the proto-inflationary patch is not a
sign of how natural a starting point it is, but of how extremely
difficult it would be to simply find ourselves there from a generic state.
If inflation is to play a role in explaining the initial conditions
of the universe, we need to understand how it arises from some
specific condition, rather than simply appealing to randomness.

\subsubsection{Reversibility version}

The reversibility version of the unitarity critique does not explicitly
use the entropy as a measure on the space of randomly chosen initial states,
but instead attempts to characterize
what we should consider a ``natural'' starting condition by implicitly
invoking the double standard principle and examining what we would
consider natural final conditions for a collapsing universe
\cite{unruh,Hollands:2002yb}.

Even in the absence of a reliable measure on the space of initial
conditions, it seems sensible to
invoke Liouville's theorem, which states that dynamical evolution
preserves
the measure on phase space.  In other words, given a measure
$\mu$ on the phase space of a system, and two subspaces $A_i$ and $B_i$
representing two different classes of initial conditions that evolve
into final conditions $A_f$ and $B_f$, we will have
\be
  {\mu(A_i)\over \mu(B_i)} = {\mu(A_f)\over \mu(B_f)}\ .
\ee
If $A_i$ represents initial conditions that are ``unlikely'' and $B_i$
represents initial conditions that are ``likely,'' they evolve into
final conditions $A_f$ and $B_f$ that are unlikely and likely,
respectively.

Assuming Liouville's theorem holds, the reversibility critique is
simply the statement that, within the set of initial conditions that
evolve into a universe like ours, the subset that does not pass through
an inflationary phase is much larger than the one that does.  It is
easiest to see this by starting with our present universe considered in
a coarse-grained sense ({\it i.e.}, the space of all microstates that
macroscopically resemble our current universe).  If we take states in
this set and evolve them backwards in time, or equivalently consider
collapsing universes with conditions otherwise similar to our own, we
generically expect an increasing departure from homogeneity as space
contracts and gravitational perturbations grow.  Black holes will
tend to form, and the approach to a singularity will be highly
non-uniform, little resembling the smooth Big Bang of our actual past
history.  Certainly it seems extremely unlikely that the universe
would smooth out and ``anti-inflate,'' with a period of re-freezing
in which hot matter cleverly assembles itself into a smooth inflaton
field which then rolls up its potential into a contracting quasi-de~Sitter
phase.

The reversibility critique is thus very simple: the measure on the
space of conditions that anti-inflate in a contracting universe is a
negligible fraction of that on the space that collapse inhomogeneously
(or even relatively homogeneously, but without anti-inflation);
but the ratio of these measures is the same as the ratio of the measure
on the space of inflationary initial conditions to that on the space
of all conditions that evolve into a universe macroscopically like
ours. Therefore, if our initial conditions are truly chosen randomly,
inflationary conditions are much less likely to be chosen than some
other conditions that evolve into our universe (which are still a
tiny fraction of all possible initial conditions).  In other words,
nothing is really gained in terms of naturalness by invoking an
early period of inflation; as far as random initial conditions go,
it requires much less fine-tuning to simply put the universe in some
state that can evolve into our present conditions.

The idea that our current universe is more likely to be randomly chosen
than a small, smooth patch dominated by a large potential energy seems
intuitively nonsensical.  Our current universe is large, complicated,
and filled with particles, while the proto-inflationary patch is tiny,
simple, and practically empty.  But our intuition has been trained
in circumstances where the volume, or particle number, or total energy
is typically kept fixed,
and none of these is true in the context of quantum field theory coupled
to gravity.

\subsubsection{Causal-patch version}
\label{causalpatch}

An interesting perspective on the likelihood of different initial
conditions for the universe arises from the causal-patch description
of spacetime 
\cite{Banks:2000fe,Banks:2001yp,Dyson:2002nt,Susskind:2002ri}.\footnote{An
alternative implementation of causal-patch
physics is the ``holographic cosmology'' of Banks and Fischler
\cite{Banks:2001px,Banks:2003ta,Banks:2004vg}.  This scenario offers
a very different perspective on the dynamics of the early universe,
which we do not consider in this paper.}
This view derives from two profound lessons of
black-hole physics:  the holographic principle
\cite{'tHooft:1993gx,Susskind:1994vu,Bousso:2002ju} and black-hole
complementarity \cite{Susskind:1993if}.  A concrete consequence
of the holographic principle is Bousso's covariant entropy bound,
which places a limit on the entropy that can be contained within a
region \cite{Bousso:1999xy}.
More specifically, given a spacelike region $\Sigma$ with
boundary $\partial \Sigma$, the amount of entropy that can pass through an
appropriately constructed inward-pointing null surface emanating from the
boundary is given by the area of $\partial\Sigma$ as measured in Planck
units.  Black-hole complementarity, meanwhile, suggests that consistent
descriptions of physical situations may only include events within
the horizon of a single observer.  In the context of a de~Sitter
spacetime, these two ideas lead to the causal patch perspective, which
states that we should only consider the physics inside a single
causal patch bounded by the de~Sitter horizon.  Since the area of
such a horizon is finite (with radius $R_{\rm dS} = \sqrt{3/\Lambda}$),
the causal patch has only a finite number of degrees of freedom.

Once we imagine that there are only a finite number of degrees of
freedom, the lessons of Poincar\'e recurrence discussed
in Section~\ref{arrow} become relevant.  There are only a finite
number of configurations of the causal patch, and over an infinite
period of time every possibility will be sampled infinitely often.
The equilibrium configuration is empty de~Sitter, but universes like
our own can arise through statistical fluctuations, either directly
or via a period of inflation.  Unfortunately,
as discussed above, it is exponentially easier to fluctuate
into a universe which is
anthropically allowed but thermodynamically nonsensical than into a
universe that evolved normally from an extremely low-entropy initial
state.

Dyson, Kleban, and Susskind \cite{Dyson:2002pf} have forcefully
emphasized this apparent contradiction between causal-patch physics
and thermodynamics.  A possible reconciliation
has been suggested by Albrecht and Sorbo \cite{Albrecht:2004ke}.
They argue that inflation can be favored over ordinary evolution,
even if there are only a finite number of degrees of freedom,
by calculating the entropy of the proto-inflationary patch to
include the entropy of the larger de~Sitter region in which it
is embedded.  They suggest that this preference for inflation
can resolve the ``Boltzmann's Brain'' paradox, since inflation
leads to a large-volume universe with many brains.  There is an
apparent tension, however, between this point of view and the
reversibility critique of inflation.  In a state of thermal equilibrium,
any one kind of fluctuation should be exactly equally as likely as
its time-reversed counterpart.  We can certainly imagine a
fluctuation in which a contracting universe is gradually assembled,
smoothes itself out, and anti-inflates; but for the same reasons given
in the previous subsection, this seems much less likely than
contraction to a highly inhomogeneous Big Crunch.  It seems very
difficult to derive a true arrow of time from fluctuations about
an equilibrium background.

Our perspective is quite different, in that we imagine there are
truly an infinite number of degrees of freedom, and that we may
sensibly speak of causally disconnected parts of the universe.
For us, de~Sitter is not an equilibrium state about which we are
fluctuating, but a metastable state that eventually decays via
spontaneous inflation, as discussed below.
Strictly speaking, these assumptions are not inconsistent with the
covariant entropy bound.  The initial proto-inflationary patch must
be somewhat larger than the inflationary Hubble radius; if such
a spacetime remained de~Sitter for all time, the corresponding
light-sheets emanating inward from the boundary would not form a closed
surface, and the holographic bound places no constraint on the number
of degrees of freedom inside.  One might attempt to subdivide the patch
into smaller regions and calculate the bound for each sub-region, but
that would only make sense if the bound were additive for nearby
regions ({\it i.e.}, if the entropy flux were localized into individual
regions).  In other words, such a subdivision neglects the entropy
corresponding to the relative configurations of neighboring patches,
which can presumably be important.

Of course, the proto-inflationary patch does not remain in a
quasi-de~Sitter phase forever, but eventually reheats and the Hubble
parameter begins to decrease.  In a spatially flat universe with zero
cosmological constant, the light-sheets would ultimately close and the
holographic bound would be finite.  It is therefore amusing to note
that the current acceleration of the universe implies that we are once
again entering a de~Sitter phase, and the light-sheets emanating from
the boundary of the original proto-inflationary patch lie outside our
de~Sitter horizon.  Thus, they will again not form a closed surface
(although the situation can change if the dark energy is quintessence rather
than a cosmological constant \cite{Hellerman:2001yi,Fischler:2001yj}).
We do not know whether there is some deep connection between inflation,
holography, and the current value of the vacuum energy.  For the rest
of this paper, in any event, we will simply assume that a semiclassical
treatment with an infinite number of degrees of freedom is valid.

\subsection{Unitarity and Degrees of Freedom}

In any version of this critique of inflation, there is a
crucially important assumption:  that the evolution of the comoving
volume from the early universe to today is unitary (and hence reversible).
The assumption of
unitarity seems innocuous, but is extremely profound in this context; it
underlies the claim that the small proto-inflationary patch and our
current universe are
the same system, just in two different configurations.

An obvious prerequisite for unitary evolution is the conservation of
the set of degrees of freedom characterizing the system.  Such an assumption
is so deeply ingrained in our understanding of dynamics that it is not
usually spelled out explicitly, but in the case of an expanding spacetime
the issue is not so clear.  Indeed, unitarity is directly at odds with
the reasoning behind the idea that it is much easier to begin inflation than
to simply find the universe in its current state:  the assertion that
there are not that many degrees of freedom
that need to be in their vacuum state in order for a small region to
begin inflating.  If the evolution is truly unitary,
there are a fantastically large number of
such degrees of freedom, most of which would correspond in a Fourier
decomposition to modes with frequencies
well above the Planck scale, and each of these degrees of freedom
needs to be separately put into its vacuum state.  These modes would
generically be excited in a time-reversed or collapsing universe, as
inhomogeneities grew and black holes were formed. In other words, the
idea that degrees of freedom are neither created nor destroyed
dramatically undercuts the hope that it is not too unlikely to find
a patch of the universe in an appropriately proto-inflationary
configuration.

It is by no means obvious that the evolution should be unitary.
A well-known idea of Hawking (recently recanted)
suggests that unitarity is violated in
the process of black-hole evaporation \cite{Hawking:1976ra};
it is equally reasonable
to imagine that other processes involving semiclassical quantum
gravity are also non-unitary, including inflationary cosmology.  
Kofman, Linde, and Mukhanov \cite{Kofman:2002cj} have pointed out
that there is an apparent violation of unitarity in the process
of reheating, due to particle production.  Whether one chooses to 
think of this process as truly non-unitary is to some extent a matter of
choice, depending on one's point of view toward collapse of the
wavefunction; from a Copenhagen point of view it is truly non-unitary,
while from a many-worlds perspective the evolution of the entire
wavefunction is perfectly unitary.  (See also the response by Hollands
and Wald \cite{Hollands:2002xi}, and the paper by Mathur \cite{Mathur:2003ez}.)
If unitarity is violated, degrees of freedom
are brought into existence as the universe expands, so that our current
universe has a much larger number of degrees of freedom than were
present in the corresponding comoving volume at the beginning of
inflation.
A violation of unitarity is a dramatic possibility, but
it could conceivably vitiate the critique of inflation just
discussed; if there are very few degrees of freedom initially in the
system, the initial state may have an entropy that is nearly maximal
even if it is much lower than the entropy today.

Our perspective in this paper will be to maintain unitarity, but
instead examine carefully the notion of a generic state in a theory
with gravity.  If the beginning of inflation relies on a random
fluctuation from chaotic initial conditions, it is unclear what is
gained by invoking a fluctuation to a configuration with such an
unnecessarily low entropy.  In the next section we will argue that the
conventional picture of ``generic initial conditions'' is misleading.
Rather than a state with large fluctuations at or near the Planck
scale, we will suggest that a generic state is close to empty
space.  The entropy density of such a configuration will be very low,
but the total entropy will be very large.  {}From this specific kind of
configuration, in the presence of a positive vacuum energy, quantum
fluctuations of an inflaton field can lead to the spontaneous onset
of inflation, which then proceeds normally, eventually reheating and
creating regions similar to our observed universe.

\section{Typical States in Theories with Gravity}
\label{typicalstates}

\begin{figure}[t]
\vskip-0cm
\centerline{\psfig{figure=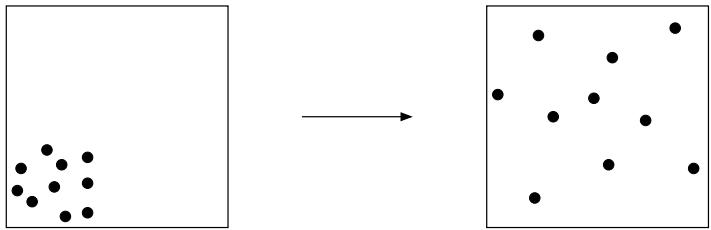,height=3cm}}
\caption{In the absence of gravity, a box of gas evolves to an
equilibrium state with a homogeneous distribution.}
\label{box1}
\end{figure}

A textbook example of the approach
to thermal equilibrium from a low-entropy state is the evolution of a gas
of particles in a box, as portrayed in Figure~\ref{box1}.  If the gas
is originally located in some small corner of the box, it will typically
evolve toward a higher-entropy state by spreading throughout the box.
The state of maximum entropy will represent thermal equilibrium, in which
the coarse-grained state (described in terms of macroscopic observables
rather than the microscopic information about each molecule of the gas)
remains static.

\begin{figure}[t]
\vskip-0cm
\centerline{\psfig{figure=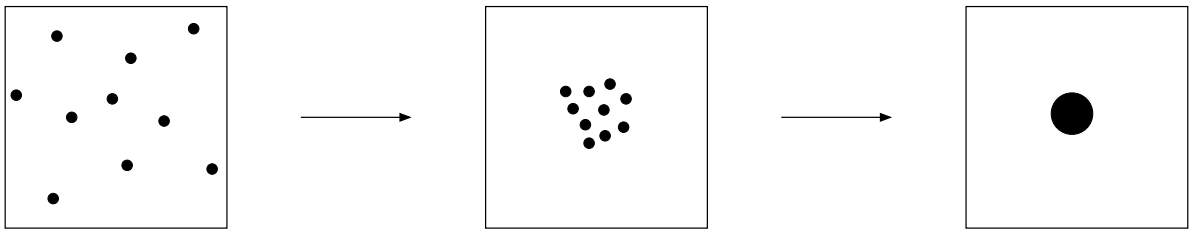,height=3cm}}
\caption{In the presence of gravity, a box of gas with size greater than
the Jeans length evolves to form
a black hole via gravitational instability.}
\label{box2}
\end{figure}

As Penrose has emphasized \cite{penrose}, this story is different once we turn
on gravity.  (In this section we imagine that the vacuum energy is zero,
so that empty space is flat.)
If the size of the box is greater than the Jeans length
\be
  \lambda_J = \sqrt{{\pi v_s^2 \over G\rho}}\ ,
\ee
where $v_s$ is the sound speed and $\rho$ the energy density,
the gas will be unstable to gravitational clumping, as shown in
Figure~\ref{box2}.
The tendency will therefore {\em not} be for the gas to become
homogeneous, but rather concentrated in a localized overdense region.
Given enough time, the overdense region should form a black hole; even
if the equation of state of the gas supports a metastable compact
object (such as a planet or degenerate star), there will be some
amplitude for a quantum fluctuation to collapse the object to a black
hole.  It is therefore clear, even in the absence of a reliable theory
of gravitational entropy, that the compact objects have a greater entropy
than the homogeneous gas; whatever entropy is, it tends to increase
in the course of typical evolution from a state with non-maximal entropy.

Although the black hole is higher entropy than the original homogeneous
gas, it cannot be a maximum-entropy state, since it is not static; the
black hole will gradually evaporate by emitting Hawking
radiation.\footnote{It is often stated that ``black holes are
configurations of maximum entropy,'' but this is true only
when the area of the boundary is held fixed.  In a realistic context
where spacetime evolves and areas are not fixed, black holes do not
maximize the entropy.  (If they did, they wouldn't evolve into something
else.)  Stable radiating black holes may exist in anti-de~Sitter space,
but that is not the real world.}
Although we could imagine constructing a sturdy box with reflecting
boundary conditions in such a way that the black hole eventually comes
into equilibrium with the radiation in the box, such a configuration
is clearly contrived and unstable; any rupture in the integrity of the
box would allow the radiation to escape, and there will be some small
probability that the entire box will tunnel into the black hole.
Instead, let us take the
box itself to represent some simple set of boundary conditions, and ask
what the ultimate state of the system might be.  There are
innumerable possible conditions that could pertain outside the box,
but it is useful to consider two limiting cases: an asymptotically-flat
empty universe, and periodic boundary conditions.

\begin{figure}[t]
\vskip-0cm
\centerline{\psfig{figure=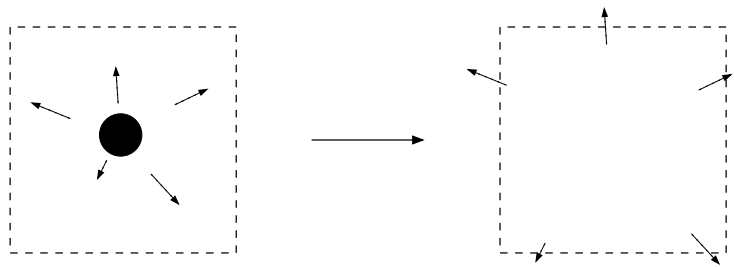,height=3.5cm}}
\caption{With asymptotically flat boundary conditions, the black
hole evaporates away, leaving a thin gas of radiation in an
increasingly empty space.}
\label{box3}
\end{figure}
If the universe is empty outside the fictitious box, it is obvious
what happens:  the black hole evaporates away, and the ultimate state
is nearly-flat empty space, as the Hawking radiation disperses to
infinity as shown in Figure~\ref{box3}.
One can prove \cite{Zurek:1985gd,Frolov:1993fy,Sorkin:1997ja,Flanagan:1999jp}
certain versions of the Generalized Second Law,
which guarantees that the radiation itself, free to escape
to infinity, does have a larger entropy than the original black hole.
(For the long-term fate of objects in an astrophysical context,
see \cite{Adams:1996xe}.)

\begin{figure}[t]
\centerline{
\psfig{figure=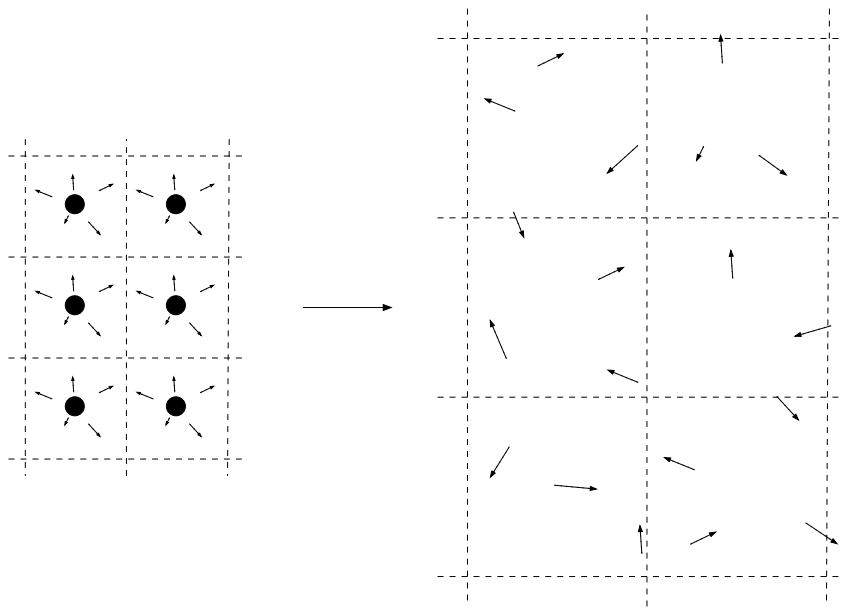,height=8cm}}
\caption{In an expanding universe, black holes evaporate and the
resulting radiation becomes increasingly rarefied, so that the universe
approaches flat spacetime.}
\label{box4}
\end{figure}

\begin{figure}[t]
\centerline{
\psfig{figure=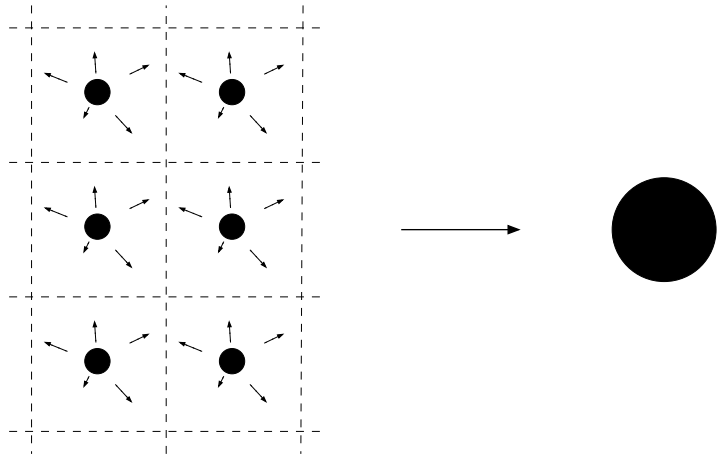,height=6cm}}
\caption{A contracting universe filled with black holes collapses
to a singularity.  If the ``universe'' is only a compact piece of a 
larger space, the singularity will be a black hole that can radiate
away.}
\label{box5}
\end{figure}

The case of periodic boundary conditions is more subtle.  Such a
configuration describes a lattice of black holes throughout space,
equivalent on large scales to a cosmological spacetime with a certain
average energy density.  Einstein's equation then tells us that the
overall scale factor of the space must be expanding or contracting.  If
it is expanding, the black holes will be able to evaporate away, leaving
a universe of gradually diluting radiation, ultimately approaching
flat spacetime, as shown in Figure~\ref{box4}.  We are therefore left with
a final state similar to the asymptotically-flat case of the previous
paragraph.

If the scale factor is contracting, however, the density will
increase until the universe hits a Big Crunch singularity in the future,
as shown in Figure~\ref{box5}.
But we can argue that collapse to a cosmological singularity is
a non-generic situation.  More specifically, specifying initial data
in any region with compact support is never sufficient to guarantee
a future cosmological singularity.  (We are assuming that the universe
is non-compact; if spatial sections are compact, Big Crunch singularities
can be generic.)  Given initial data in some local
region describing a collapsing universe, we can always surround the
region by a large but finite region of empty space.  Now the
collapse actually describes a black hole, which eventually radiates
away as before.  To ensure that the entire universe collapses to a
final singularity requires that we specify initial data over the
entire infinite spacelike hypersurface; we therefore conclude that such
behavior is non-generic, and evolution to flat empty space is the only
robust outcome.  (This conclusion is not
absolutely necessary for our general picture; if the argument against
future singularities is unconvincing, simply replace ``empty space''
in subsequent claims by ``empty space or a future singularity.'')

The examples considered in this section provide anecdotal evidence for
a straightforward claim:  in a theory with gravity (and vanishing
vacuum energy), the closest thing to a maximal-entropy,
thermal-equilibrium
state is flat empty space.  Another way to reach the same conclusion is
to simply take any configuration defined on a spacelike hypersurface, and
realize that we can increase the entropy by taking the same set of
excitations (whether in matter fields or gravitational waves) and
spread them apart by increasing the scale factor, thereby increasing the
allowed phase space for each particle.  This expansion doesn't violate
any conservation laws of the system, so there is no obstacle to
configurations eventually increasing their entropy in this fashion.

Consequently, there is no reason to expect randomly-chosen or generic
conditions to feature large curvatures and Planck-scale fluctuations.
According to everything we know about gravity, large curvatures are
entropically disfavored, tending to ultimately smooth themselves out
under ordinary evolution.  This is a direct consequence of the ability
of a curved spacetime to evolve towards perpetually
higher entropy by having the universe expand, unlike gas trapped in a box.
{}From this point of view, it should not be considered surprising that
we live in a relatively cold, low-curvature universe; the surprise
is rather that there is any observable matter at all, much less
evolution from an extremely hot and dense Big Bang.

The idea that high-entropy states correspond to nearly-flat empty
space makes the conundrum of the initial conditions for inflation
seem even more acute -- such a condition seems very different from
the chaotic initial conditions often invoked in discussions of inflation.
In the next section we explore how inflation
can originate from quantum fluctuations in a nearly-empty universe
with a small positive cosmological constant and an appropriate inflaton
field.

\section{Spontaneous Inflation from Cold de Sitter Space}

In the previous section we argued that generic initial conditions in
a theory with gravity tend to evolve to flat empty space, which is
correspondingly the highest-entropy state in the theory.  (The entropy
{\it density} is low, but it is the total entropy which tends to increase
according to the Second Law.)  If initial conditions for the universe
are randomly chosen, with high-entropy states correspondingly
preferred, the natural question is then why we observe any matter in
the universe at all, much less an extremely dense Big Bang.

A way out of this conundrum is possible if ``empty space'' is not a
perfectly stable state, but rather is subject to instabilities that
can produce universes like our own.  A mechanism for such an instability
may be provided by the small positive cosmological constant that has
apparently been discovered
\cite{Riess:1998cb,Perlmutter:1998np,Carroll:2000fy}.  (If the acceleration
of the universe is due to dynamical dark energy or a modification of
gravity on large scales, the discussion to follow would change in
important ways and perhaps become irrelevant.)

The basic idea is very simple.  In the presence of a positive vacuum
energy, it will remain true that most states tend to empty out to empty
space, but ``empty space'' will correspond to de~Sitter rather than
Minkowski spacetime.  Unlike Minkowski, which has zero temperature, and
a de~Sitter space with vacuum energy
density $\rho_{\rm vac} = M_{\rm vac}^4$ will have a Gibbons-Hawking
temperature
\be
  T_{\rm dS} = {H \over 2\pi} \sim {M_{\rm vac}^2 \over \mpl}\ .
  \label{ght}
\ee
This temperature gives rise to thermal fluctuations in any fields in
the theory.  In this section, we describe how fluctuations in an
appropriate inflaton field $\phi$ can lead to the spontaneous onset of
inflation, which can then continue forever as in the standard story
of eternal inflation.  This idea is not new; Garriga and Vilenkin,
for example, have proposed that thermal fluctuations can induce
tunneling from a true de~Sitter vacuum to a false vacuum at higher
energies, thus inducing spontaneous inflation \cite{Garriga:1997ef}.
(There is also a body of literature that addresses the creation of
inflating universes via quantum tunneling, either ``from nothing,''
at finite temperature, or from a small patch of false vacuum
\cite{Vilenkin:1982de,Vilenkin:1983xq,Linde:1983mx,Vilenkin:1984wp,
Starobinsky:1986fx,Goncharov:1986ua,
Farhi:1986ty,Vilenkin:1987kf,Farhi:1989yr,Fischler:1989se,Fischler:1990pk,
Linde:1991sk}.)  In our discussion is that we examine the case of an 
harmonic oscillator
potential without any false vacua; in such a potential we can simply
fluctuate up without any tunneling.  The resulting period of inflation
can then end via conventional slow-roll, which is more phenomenologically
acceptable than tunneling from a false vacuum (as in ``old inflation''
\cite{Guth:1980zm}).  Thus, the emptying-out of the universe under
typical evolution of a generic state can actually provide
appropriate initial conditions for the onset of inflation, which then
leads to regions that look like our universe.

We should emphasize that our calculation ignores many important
subtleties, most obviously the back-reaction of the metric on the
fluctuating scalar field.  Nevertheless, our goal is to be as 
conservative as possible, given the limited state of our current
understanding of quantum gravity.  In particular, it is quite possible
that a similar tunneling into inflation may occur even in a
Minkowski background (see {\it e.g.} \cite{Linde:1991sk}).  In
our calculation we simply discard the vacuum fluctuations that are
present in Minkowski, and examine only the additional contributions
from the nonzero de~Sitter temperature.  We believe that the resulting
number (which is fantastically small) provides a sensible minimum
value for the probability to fluctuate up into inflation.  The true
answer may very well be bigger; for our purposes, the numerical
result is much less important than the simple fact that the background
is unstable to the onset of spontaneous inflation.  Clearly this
issue deserves further study.

\subsection{Eternal Inflation}

\begin{figure}[t]
\centerline{
\psfig{figure=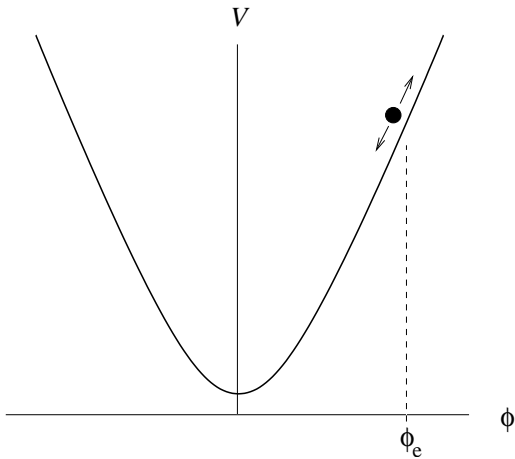,height=6cm}}
\caption{Quantum fluctuations up a potential drive eternal inflation.}
\label{eternalfig}
\end{figure}

We first recall the basics of eternal inflation 
\cite{Vilenkin:1983xq,Linde:1986fc,
Linde:1986fd,Goncharov:1987ir}.  As in
Section~\ref{critique}, consider a massive scalar field with an
potential $V(\phi)={1\over 2}m^2\phi^2+ V_0$ and a mass
$m\sim 10^{13}$~GeV.  (This example is merely illustrative; the details
of the potential are not crucial, so long as some simple requirements
are met.)  Classically, the field will roll down the
potential according to
\be
  \ddot\phi + 3H\dot\phi + m^2\phi = 0\ .
\ee
In the slow-roll regime, where we can ignore the $\ddot\phi$ term in
this equation and the energy density is dominated by ${1\over 2}m^2\phi^2$,
the Friedmann equation implies
\be
  H \approx \sqrt{4\pi\over 3}{m\over \mpl}\phi\ ,
\ee
and the classical motion of the field obeys
\be
  \dot\phi \approx {m\mpl\over \sqrt{12\pi}}\ .
\ee
However, superimposed on this
classical motion are quantum fluctuations in the field, as shown in
Figure~\ref{eternalfig}.
On Hubble-radius scales these fluctuations are given by the Gibbons-Hawking
temperature (\ref{ght}),
\be
  \delta\phi = {H\over 2\pi}\ .
\ee
These quantum fluctuations become comparable to the classical motion
when
\be
  \delta\phi \approx {\dot\phi\over H}\ ,
\ee
or for any field value $\phi \geq \phi_{\et}$, where
\be
  \phi_{\et}^2 = \sqrt{\frac{3}{16 \pi}}\frac{\mpl^3}{m}\ .
  \label{n}
\ee
For $m\sim 10^{13}$~GeV, this implies $\phi_{\et} \sim 10^3\mpl$, with
a corresponding Hubble parameter
\be
  H_{\et} \sim \sqrt{m\mpl}.
\ee
[Note that these are the values for which inflation is eternal; the
values (\ref{phii}) and (\ref{hi}) are the less-stringent requirements
for ordinary inflation.]
If this point is reached, inflation never truly ends.  After each
$e$-fold of expansion, the field is likely to roll {\it up} the
potential in some region of space, while continuing to roll down
in most of the comoving volume.  However, in the region where the
field rolls up, the increased energy density leads to a faster expansion
rate and consequently a greater increase in the volume; it's
straightforward to show that the physical volume of inflating space
continues to grow indefinitely.  Although the field along
any given geodesic is likely to fall down the potential and reheat
as usual, inflation continues to occur somewhere else in the universe.

\subsection{Fluctuating into Inflation}

The idea we explore in this section is derived directly from the
philosophy of eternal inflation, but we begin by imagining that the
field is in its vacuum state at the bottom of a potential in a de~Sitter
spacetime.  Given the nonzero Gibbons-Hawking temperature, there will
necessarily be fluctuations in the scalar field over and above the
zero-point fluctuations expected in a Minkowski background.  There is
therefore some chance that the field can spontaneously
fluctuate all the way up to $\phi_{\et}$ in a patch of radius $H_{\et}^{-1}$,
thereby providing an appropriate initial condition for the onset of
inflation.\footnote{From the point of view of a hypothetical
observer outside the fluctuation, the process of spontaneous inflation
presumably looks like the creation of a small black hole
\cite{Farhi:1986ty}.  Furthermore, we are assuming that the Hubble parameter
in the fluctuating region is positive; half of the time it will be
negative, leading to a rapid collapse to a singularity inside a
black hole.}  The probability is of course small, but if we have
infinitely long to wait it will eventually occur.

On scales much smaller than the de Sitter radius, spacetime looks flat
and we can use finite temperature field theory in flat space to obtain
an order of magnitude estimate for the probability that eternal inflation
will start.
Suppose that at some point in spacetime the
inflaton is in its ground state, with a Gaussian wavefunction.
The probability density of finding the field at some value
$\phi(x)$ is the wavefunction squared:
\be
{\cal P}[\phi(x), T] = N_{\cal P} \exp
\left( -\frac{\phi^2(x)}{2\left< \phi^2(x) \right>} ,
\right),
\label{p}
\ee
where the normalization is
\be
N_{\cal P} = \frac{1}{\sqrt{2 \pi \left< \phi^2(x) \right>}}.
\ee
The variance $\left< \phi^2(x) \right>$ is obtained by evaluating the
finite temperature correlation function \cite{iz}
\be
\left<\phi(x), \phi(x') \right>
= \int \frac{d^3k}{(2\pi)^3}\frac{1}{2E_{k}}\left(e^{-ik(x - x')} \left(1
+ f(E_{k}) \right) + e^{ik (x - x')} f(E_{k}) \right)
\label{c2}
\ee
at the same point.

However, we need to first average the fluctuations in the field over some
scale with a smoothing function $f$ to properly study fluctuations on that
scale.  We therefore define the smoothed field operator,
\be
\bar{\phi}(\vx) =\int \phi(\vy) g(\vx - \vy) d^3 y .
\ee
The appropriate scale here is the inflationary Hubble scale.  As a
smearing function, we will use a Gaussian of width $H_{\et}^{-1}$,
\be
g(\vx) = N_g e^{-\vx^2/2 H_{\et}^{-2}} ,
\ee
where the appropriate normalization factor is
\be
N_g = \left(\frac{H_{\et}}{\sqrt{2\pi}}\right)^3 .
\ee
The answer we obtain will thus correspond to the probability of the
field jumping up the potential over one proto-inflationary Hubble
volume (much smaller than the Hubble volume of the ambient de~Sitter
space in which the fluctuation occurs).

Averaging the fields in the correlator (\ref{c2}) at equal times, we get
\be
\left<\bar{\phi}(\vx), \bar{\phi}(\vx') \right>
= N_g^2 \int d^3y d^3y'
\frac{d^3k}{(2\pi)^3}\frac{1}{2E_{k}}e^{-(\vy-\vx)^2H_{\et}^2/2}
e^{-(\vy'-\vx')^2H_{\et}^2/2}\left(e^{i\vk\cdot(\vy - \vy')} \left(1 + f \right) +
e^{-i\vk\cdot(\vy - \vy')} f\right).
\label{c3}
\ee
Integrating over $\vy$ and $\vy'$ and setting $\vx = \vx'$, we obtain
\be
\left<\bar\phi^2(x)\right>
= \int \frac{d^3k}{(2\pi)^3}\frac{1}{2E_{k}} e^{-3k^2/H_{\et}^2}
\left(1 + 2 f(E_{k}) \right).
\label{c1}
\ee
For bosons, the phase space density is the Bose-Einstein distribution,
\be
f(E_{k}) = \frac{1}{e^{E_{k}/T_{\rm dS}} - 1}.
\ee
The $f$-independent piece of the correlation function (\ref{c1})
is the usual vacuum-fluctuation
divergence present in Minkowski spacetime at zero temperature;
we will renormalize this to zero under the assumption that Minkowski
spacetime should be stable.  The remaining temperature-dependent
contribution is a finite integral,
\be
\left<\bar\phi^2(x)\right>_{\ren}
= \int \frac{d^3k}{(2\pi)^3}\frac{1}{E_{k}} e^{-3k^2/H_{\et}^2} f(E_{k}).
\label{c}
\ee
By changing variables from $|\vk|$ to $E$, we get
\be
\left<\bar\phi^2(x)\right>_{\ren}
= \frac{1}{2\pi^2}e^{3m^2/H_{\et}^2}\int_m^\infty dE
\sqrt{E^2-m^2} e^{-3E^2/H_{\et}^2}\frac{1}{e^{E_{k}/T_{\rm dS}}-1},
\ee
where $m$ is the mass of the scalar, which we will take to be $\sim
10^{13}$~GeV.  Given that $\rho_{\Lambda} \sim 10^{-47}$~GeV$^4$,
$T_{\rm dS}$ is
\be
T_{\rm dS} = \frac{H_{\rm dS}}{2\pi} = \sqrt{\frac{2G}{3\pi}\rho_{\Lambda}} \sim
10^{-43} \ {\rm GeV}.
\ee
Because $m \gg T$ (we will
henceforth drop the subscript on $T$), and since $E$ is
always larger than $m$, we can safely assume $e^{E_{k}/T} \gg 1$.
Changing variables again to $y \equiv E/m$, we have
\be
\left<\bar\phi^2(x)\right>_{\ren}
= \frac{m^2}{2\pi^2} e^{3m^2/H_{\et}^2} \int_1^\infty dy
\sqrt{y^2-1}e^{-my/T} e^{-3m^2y^2/H_{\et}^2}.
\label{}
\ee

This has no closed form solution.  However, the factor
$e^{-3m^2y^2/H_{\et}^2}$ is close to unity over the interval where the
other factors are non-negligible, so that
it can be safely set to unity.  The resulting integral,
\be
\left<\bar\phi^2(x)\right>_{\ren}
= \frac{m^2}{2\pi^2} e^{3m^2/H_{\et}^2} \int_1^\infty dy
\sqrt{y^2-1}e^{-my/T},
\label{1}
\ee
can be evaluated in terms of a modified Bessel function, $K_1$, as
\be
\left<\bar\phi^2(x)\right>_{\ren} = \frac{mT}{2\pi^2}K_1(m/T)e^{3m^2/H_{\et}^2} .
\ee
For large $m/T \ (\sim 10^{56})$, $K_1$ is approximately
\be
K_1(m/T) \sim \sqrt{\frac{\pi T}{2m}} e^{-m/T} .
\ee
Finally we find the variance to be
\be
\left<\bar\phi^2(x)\right>_{\ren} = \frac{T}{2\pi}\sqrt{\frac{mT}{2
\pi}}e^{-m/T}e^{3m^2/H_{\et}^2}.
\label{d}
\ee

Eternal inflation begins whenever $\phi$ fluctuates to values
larger than $\phi_{\et}$ (or less than $-\phi_{\et}$).  Therefore, the
probability per spacetime volume $H_{\et}^{-4}$
that the scalar field spontaneously fluctuates
sufficiently far up its potential to induce eternal inflation is
\be
P(\phi_{\et}) = 2\int_{\phi_{\et}}^{\infty} {\cal P}[\phi, T] d\phi.
\ee
Putting in (\ref{p}), we obtain
\be
P(\phi_{\et}) = 1 - \erf\left( \frac{\phi_{\et}}{\sqrt{2\left<\phi^2\right>_{\ren}}}
\right).
\ee
The error function $\erf(x)$ can be approximated
for large $x$  by
\be
\erf(x) \approx 1 - \frac{e^{-x^2}}{x \sqrt{\pi}}.
\ee
This results in
\be
P(\phi_{\et}) = \frac{\exp\left( -\frac{\phi_{\et}^2(x)}{2\left< \bar\phi^2(x)
\right>_{\ren}}\right)}
{{\sqrt{\pi}\frac{\phi_{\et}}{\sqrt{2\left<\bar\phi^2\right>_{\ren}}}}}.
\label{prob}
\ee
Substituting (\ref{n}) and (\ref{d}) into (\ref{prob}), we find that the
probability for the spontaneous onset of eternal inflation in a
spacetime volume of radius $H_{\et}^{-1}$ in an ambient de~Sitter spacetime
with temperature $T$ to be
\be
P = \frac{\exp\left(-\pi\frac{\mpl^3}{m T}
  \sqrt{\frac{6}{mT}}e^{m/T}e^{-3m/\mpl}\right)}
  {\pi \sqrt{\frac{\mpl^3}{m T}
  \sqrt{\frac{6}{mT}}e^{m/T}e^{-3m/\mpl}}}.
\ee
Evaluating this mess for
$m \sim 10^{13}$ GeV, $\mpl \sim 10^{19}$ GeV, and $T \sim
10^{-43}$ GeV, we obtain
\be
  P \sim 10^{-10^{10^{56}}},
  \label{number}
\ee
an appropriately tiny answer.\footnote{We suspect that this may be
smallest positive number in the history of physics, but we haven't
done an exhaustive search to check.}

The important feature of this probability, calculated in the context
of a specific model, is not its actual numerical value, but simply
the fact that it is nonzero (which is certainly not a surprise,
given our assumptions).  The context in which we have performed the
calculation -- calculating renormalized fluctuations of a scalar field
in a fixed de~Sitter background, and then imagining that gravitational
back-reaction leads to the onset of inflation -- is by no means
well-understood, although no less so than most discussions of eternal
inflation.  The crucial point is that it is quite natural for de~Sitter
to be unstable to the onset of inflation, as has also been argued
elsewhere \cite{Linde:1991sk,Garriga:1997ef}.  (The instability
studied here is of course different from a possible infrared
instability of de~Sitter space; see e.g. \cite{Tsamis:1996qq}.)

\subsection{The Ultra-Large-Scale Structure of the Universe}

Given that inflation can spontaneously begin, an important remaining
question is
whether it is more likely than a fluctuation into a conventional
Big-Bang universe.  As discussed in Section~\ref{critique}, an
appropriate Robertson-Walker universe has a much higher entropy than
a proto-inflationary patch, so we might expect the
inflationary universe to be correspondingly less likely to be
spontaneously created.  However, the scenario we are describing does not
imagine that our early universe is ``chosen randomly'' in some measure on
the space of initial conditions; rather, that it evolves via a fluctuation
from a very specific pre-existing state, namely empty de~Sitter.  {}From
such a starting point, it is easier for a single mode of wavelength
$H_{\et}^{-1}$ to fluctuate up its potential than for a large collection
of modes to simultaneously fluctuate into a configuration describing
a radiation-dominated Robertson-Walker universe.\footnote{It would
obviously be useful to examine this claim more quantitatively.
It is interesting to conjecture that fluctuations of the sort we consider
would favor the creation of configurations with vanishing Weyl tensor,
thus providing a dynamical basis for Penrose's Weyl Curvature
Hypothesis \cite{penrose}.}  This claim is no
more dramatic than the claim that it is more likely to find molecules
of a gas taking up only one medium-sized corner of a box than to find
them spread evenly, {\it if} we specifically look at the box almost
immediately after the gas was released from an even smaller region in
the corner.  Although the entropy of our early
inflationary state is extremely low, it is
nevertheless larger than that of the tiny comoving volume
of de~Sitter from which it arose, in perfect accord with our conventional
understanding of the Second Law.  (It is crucial to this discussion that
the Second Law demands the increase of the total entropy rather than the
entropy density, and also that is is sensible to discuss the entropy
present along surfaces larger than one Hubble radius.)

There is another, perhaps more persuasive, argument that fluctuations
into eternal inflation dominate over those into a conventional
Big Bang -- namely, that the measure on what is more likely should come
from observers in the post-fluctuation universe, rather than from
counting events in the pre-fluctuation cold de~Sitter space.  As has
been emphasized often in the eternal-inflation literature,
once eternal inflation begins it creates an infinite volume of
livable universe in the future.  Therefore, even if fluctuations
into radiation-dominated universes (or anything similar) are more
likely than fluctuations into inflation, most observers will find
themselves to be living in a post-inflationary region just because of
the infinite volume factor associated with eternal inflation.  In
practice, using this volume factor to calculate sensible probabilities
is extremely difficult at best \cite{Vilenkin:1994ua,Garcia-Bellido:1994ci,
Vilenkin:1995yd,Winitzki:1995pg,Vilenkin:1998kr,Vanchurin:1999iv,
Garriga:2001ri,Guth:2000ka}; nevertheless, if our
only purpose is to compare inflation to non-inflation, it seems
legitimate to appeal to the fecundity of eternal inflation in creating
livable regions of spacetime.

Although the probability (\ref{number}) is quite small, fluctuating
into inflation is ultimately inevitable, since the total spacetime
volume in the cold de~Sitter phase is infinite.  One might worry
that the de~Sitter solution is only metastable, as it would be in the
string landscape picture \cite{Dasgupta:1999ss,Feng:2000if,Bousso:2000xa,
Giddings:2001yu,Kachru:2003aw,Susskind:2003kw,Douglas:2003um,Ashok:2003gk,
Douglas:2004zu,Robbins:2004hx} or other theories where the
de~Sitter phase is a false vacuum liable to decay.  However, we
don't need the decay rate to be strictly zero, or even smaller than
the inflationary-fluctuation rate (\ref{number}); all we require
is that the decay rate per Hubble volume be substantially less than
unity.  In that case, just as in old inflation, the
de~Sitter vacuum will never disappear, as the phase transition
never percolates; in fact, the physical volume of spacetime in the
de~Sitter phase will continue to increase, just as in eternal inflation.
The total spacetime volume of the de~Sitter phase
is therefore infinite, and the transition
into our proto-inflationary universe is guaranteed eventually to
occur.  Indeed, it will eventually occur an infinite number of times.

We therefore have a picture in which the universe starts in some
generic state defined on a Cauchy surface, and then is allowed to
evolve.  Local inhomogeneities may collapse to form black holes,
which eventually evaporate.  The entropy of the configuration is
increased by spreading out fluctuations into an ever-larger spatial
volume, leaving us with an empty de~Sitter solution with a small
cosmological constant.  The de~Sitter phase may or may not be unstable
to decay into a state of lower vacuum energy, but the decay rate is
assumed to be sufficiently slow that the physical volume of de~Sitter
grows without bound.  Eventually, thermal fluctuations in this
background allow a scalar inflaton field to bounce sufficiently high
up its potential that eternal inflation begins with a large vacuum
energy.  Different parts of this inflating region fall down the potential,
reheating and evolving into galaxies as in the conventional picture;
elsewhere, inflation continues forever.

\begin{figure}[t]
\centerline{
\psfig{figure=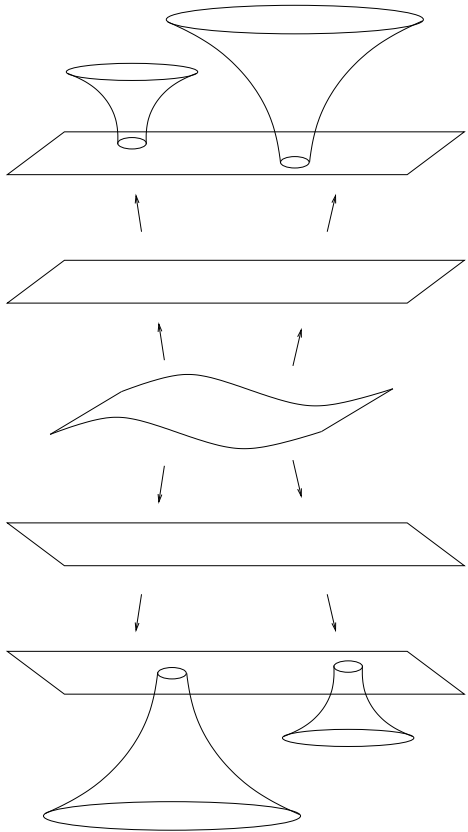,height=11cm}}
\caption{The ultra-large-scale structure of the universe.  Starting from
a generic state, it can be evolved both forward and backward in time,
as it approaches an empty de~Sitter configuration.  Eventually,
fluctuations lead to the onset of inflation in the far past and far
future of the starting slice.  The arrow of time is reversed in these
two regimes.}
\label{ultralarge}
\end{figure}

An interesting feature of this story is that, given the specification
of the conditions on the ``initial'' surface, the same set of events
will naturally occur to the {\it past} as well as to the future.
Nothing about our description involved intrinsically time-asymmetric
physics, other than the fact that the initial condition was not an
equilibrium state with maximal entropy.  According to our scenario,
this is not in any way a restriction; there is no such thing as a
state of maximal entropy, since the entropy can always increase without
bound.

We therefore have a picture of the ultra-large-scale structure of
the universe, as portrayed in Figure~\ref{ultralarge}.  Given some
generic state defined on a Cauchy surface, we evolve it to both the
past and future, and it both cases it will empty out into a cold
de~Sitter phase, after which inflation will occasionally
begin.\footnote{A well-known feature of eternal inflation is that it
does not escape
the problem of singularities, as it cannot be eternal to the past
\cite{Borde:2001nh}.  This is not really an issue for our model.
Singularities (or whatever quantum-gravitational phenomenon replaces
them in the real world) occur all the time at the center of black holes,
and eventually disappear as the black hole evaporates.  The
singularities demanded by the Borde-Guth-Vilenkin theorems need not
be spacelike boundaries for the entire spacetime.  Formally, there
are no geodesics in our spacetime along which inflation continues
at all times to the past and the future.}
This picture is statistically time-symmetric on very large scales,
and satisfies the requirements discussed in Section~\ref{arrow} for
a legitimate explanation of our observed arrow of time.  Observers
in the very far past of our universe will also detect an arrow of
time, but one that will be reversed from ours with respect to some
(completely unobservable) global time coordinate throughout the
entire spacetime.  Both sets of observers will think of the others
as living in their ``past.''

\section{Discussion}
\label{discussion}

In this paper we have addressed the question of why the entropy of
our observable universe appears to have been extremely small in the
past.  It has long been suspected that inflation might have something
to do with the answer, although the fantastically low entropy of
the proto-inflationary universe was a significant obstacle to
constructing a convincing picture.  The answer to this conundrum must
lie in the process by which inflation begins; we have proposed a
scenario in which this happens naturally from the evolution of some
arbitrarily-chosen state.  It is interesting to note that that the
recently discovered cosmological constant plays a crucial role in our
scenario for generating the arrow of time.

The basic ingredients of our picture are as follows.  We consider
an arbitrary state of the universe, specified on some Cauchy
surface.  We argued in Section~\ref{typicalstates} that the generic
evolution of a system coupled to gravity is to dilute excitations
via the expansion of spacetime.  Black holes and other inhomogeneities
may form, but they will eventually decay away, so that the
universe approaches empty space.  However, in the presence of a
positive vacuum energy and an appropriate inflaton field, the
resulting de~Sitter phase is unstable to the spontaneous onset of
inflation, instigated by the thermal fluctuations of the inflaton.
If the inflaton fluctuates sufficiently high that eternal inflation
can begin, it will continue forever, and new pocket universes will
be brought into being in those places where the field rolls down
the potential and reheats.  This chain of events happens both to the
past and the future of the specified Cauchy surface, leading to a
statistically time-symmetric universe as portrayed in
Figure~\ref{ultralarge}.  An arrow of time is dynamically generated
in both the past and the future, as the universe continually acts
to increase its entropy.

In addition to the desire to understand the origin of the arrow of
time, a primary motivation for our study has been to understand the
onset of inflation.  As discussed in Section~\ref{critique},
the unitarity critique argues that a proto-inflationary patch of
spacetime is much lower entropy than that of an ordinary
radiation-dominated universe, and hence is less likely to arise
as a random fluctuation.  In our picture, the answer to this conundrum
lies in the fact that the beginning of our observable Big Bang
cosmology does not arise as a random choice in a large phase space
of initial conditions, but rather comes via a quantum fluctuation from
a very specific prior state -- an empty de~Sitter universe that is the
natural consequence of evolution from generic initial conditions.

By taking seriously the ability of spacetime to expand and dilute
degrees of freedom, we claim to have shown how an arrow of time can
naturally arise dynamically in the course of the evolution from a
generic boundary condition.  In the classification introduced in
Section~\ref{arrow}, our proposal imagines that there do not exist
any maximum-entropy equilibrium states, but rather that the entropy
can increase from any starting configuration.  This is not, of
course, sufficient; it is also necessary to imagine that the path
to increasing the entropy naturally creates regions of spacetime
resembling our observable universe.   In the presence of a nonzero
vacuum energy and an appropriate inflaton field, we suggest that
thermal fluctuations from de~Sitter space into eternal inflation
provide precisely the correct mechanism.

A number of other cosmological scenarios have been proposed in which
the Big Bang is not a boundary to spacetime, but simply a phase
through which the universe passes.  These include the pre-Big-Bang
scenario \cite{Gasperini:2002bn,Lidsey:1999mc}, the ekpyrotic and
cyclic universe scenarios \cite{Khoury:2001zk,Khoury:2001bz,Steinhardt:2001st},
the Aguirre-Gratton scenario of eternal inflation \cite{Aguirre:2003ck},
and Bojowald's loop-quantum-gravity cosmology
\cite{Bojowald:2003cc,Bojowald:2004kt}.
To the best of our understanding, each of these proposals invokes
special low-entropy conditions on some Cauchy surface, 
either asymptotically in the far past or at some moment of minimum
size for the universe.  In our picture, on the other hand,
there is a slice of spacetime on which the entropy is minimized, but
that entropy can be arbitrarily large.
The Big Bang in our past is not a unique moment in
the history of the universe; it is simply one of the many times that
inflation spontaneously began from a background de~Sitter phase, similar
to the proposal of Garriga and Vilenkin \cite{Garriga:1997ef}.
Along with the fractal distribution of pocket universes to the far
past and far future, this feature is another reminder of the
importance of overcoming the Robertson-Walker intuition we naturally
develop by thinking about the patch of the universe we are actually
able to observe.

There are a number of points in our scenario that have yet to be
perfectly understood.  One obvious point is how the ``initial''
state is chosen.  We have argued that the ultimate evolution to the
past and future is essentially insensitive to the details of this
state, but it is nevertheless interesting to ask what it may have been.
On a more technical level, the idea of spontaneous onset of inflation
(and related ideas within the paradigm of eternal inflation) deserves
closer investigation.  We have calculated the renormalized
fluctuations of a scalar field in a fixed de~Sitter background, and
imagined that the back-reaction of the metric would lead to inflation
if the fluctuation were sufficiently large.  A more rigorous calculation
would have to involve quantum gravity from the start, at least at
a semiclassical level.
Finally, the issues concerning the number and evolution of degrees
of freedom clearly warrant further research.  Our assumption has
been that the number of degrees of freedom is infinite but fixed,
not increasing during inflation.  We therefore require a literally
infinite number of degrees of freedom to be put in their ground
states during the evolution toward de~Sitter, before the onset of
inflation.  It is important that we understand this issue more fully,
although doing so may require a deep knowledge of quantum gravity.

An interesting implication of our scenario is the non-privileged
role played by the Big Bang of our observable universe.  As in other
models of eternal inflation, the future history of spacetime takes
on a fractal structure of ever-increasing volume; in addition we
suggest that a similar structure is found in a time-reversed sense
in the far past.  The boundary conditions of the universe on
ultra-large scales are unlikely to resemble simple Robertson-Walker
geometries in any way.
Indeed, the cacophony of matter and radiation in our
observable patch of universe appears in this picture as a mere
byproduct of the relentless evolution of the larger spacetime
toward states of higher entropy.  Stars and galaxies are seen as
exaptations -- structures that have found a use other than that
for which they were originally developed by evolution.

\section*{Acknowledgments}

We would like to thank Anthony Aguirre, Andreas Albrecht,
Raphael Bousso, Brian Greene, Alan Guth, James Hartle, Stefan Hollands, 
Andrei Linde, Lorenzo Sorbo, Wati Taylor, Alex
Vilenkin, and Robert Wald for
useful conversations (even if some of them remain skeptical of our
conclusions).  We acknowledge the hospitality of the Kavli
Institute for Theoretical Physics, where some of this work was done.
This work was supported in part by the
U.S. Dept. of Energy, the National Science Foundation,
the NDSEG Fellowship, and the David
and Lucile Packard Foundation.  The KICP is an NSF Physics Frontier
Center.


\end{document}